\documentclass[aps,showpacs]{revtex4}
\usepackage{fullpage,amsfonts,amssymb,pst-node,epsfig}

\begin{document}
\title{A simple plan for the $\Delta$ resonance}
\author{N.E. Ligterink}
\affiliation{Department of Physics and Astronomy, University of
Pittsburgh, \\
3941 O'Hara Street, Pittsburgh, PA 15260, U.S.A.}
\begin{abstract}
We construct the $\Delta$ resonance as a superposition of a bare
$\Delta$ state and the $\pi N$ conti\-nu\-um. It is parametrized by
three coupling constants for local $\pi N \Delta$ and $\pi \pi N N$
couplings and the $\Delta$ mass. The latter incorporates the
mass renormalization due to the $\pi N \Delta$ interaction, 
while the results depend only weakly, if at all, on its wave-function 
renormalization. Three more renormalization constants are needed
for the derivative contact interaction. They allow one to generate the $\Delta$
resonance dynamically. A large number of fits test the quality of 
different model assumptions in the $P_{33}$, $P_{31}$ and $P_{13}$ $p$-wave 
$\pi N$ scattering channels.
\end{abstract}
\pacs{24.30.-v,
24.10.Eq,
11.80.Gw,
11.55.-m
}
\maketitle

\section{Introduction}

The $\Delta$ resonance is present in most of baryonic physics,
due to its low energy and its strong coupling to the pion-nucleon
systems.~\cite{PJ90,SL96,HDT96,P01,VDL,Arndt,Surya} 
Therefore it is is crucial to
have good and simple methods to handle the $\Delta$ resonance
and its effects on other hadronic physics. Such method should be based 
on a local theory where the renormalization is handled properly, and
divergences can be re-absorbed consistently in the interactions.
Otherwise, the method is nothing more than a particular fit of the data,
where the $\pi N \Delta$ interactions have no universal meaning.
Especially this interaction should be handled properly, due to
strength, which makes the application of perturbation theory 
dubious.
In this paper we propose such a method with local interactions,
which can be solved in closed form, without resorting to 
perturbation theory.

The $\Delta$ resonance is a spin 3/2 state, which is hard to handle in
field theory, as it is generally part of a non-renormalizable field
theory, due to the high-order derivatives in the coupling and
the propagator, which yield high-order divergences in the 
self-energy corrections. Furthermore, there are at least two other
issues that blur our image of the $\Delta$ resonance. First,
the propagator has some unknown off-shell dependence, reflected
by a free parameter. Second,
field transformations allow different representations for 
the interactions and the propagators which, however, lead to identical
observable results.~\cite{KLF00}

In this short paper all the problems raised above are circumvented.
The propagator and the interaction are lumped together, so
there are no issues concerning field transformations that
shift terms around. Furthermore, dimensional analysis, combined
with local field theory, implies that the interaction depends, in
lowest order, on
four parameters only: two $\pi N \Delta$ couplings, one $\pi \pi N N$ 
couplings, and the $\Delta$ mass. The finite 
renormalization constants are fixed by the $\Delta$ mass and the
wave function renormalization.

\section{Theory}

\begin{figure}
\centerline{\includegraphics[width=5cm]{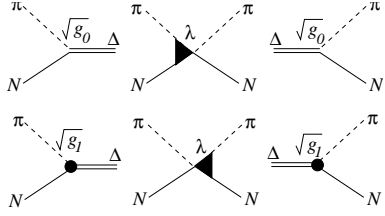}}
\caption{The four types of interactions of the Hamiltonian.
The dot represents a derivative coupling. The triangle indicates
the pion line on which the derivative acts.} \label{fig1}
\end{figure}

The self-energy diagram~\cite{Kor97} of
the $\pi N$ loop correction to the $\Delta$, which has an imaginary
part because of the $\pi N$ decay, has only two 
types of local momentum dependence of the loop integral in the rest frame:
\begin{equation}
\Sigma(E) =\frac{1}{4 \pi} \int \frac{d^3 {\bf p}}{\omega_N \omega_\pi} 
\frac{ [1;{\bf p}^2]}{E - \omega_\pi - \omega_N} \ \ ,
\label{self}
\end{equation}
due to the different terms in the traces over polarizations and spins.
The attention focusses on the real part. The same diagram occurs in
other spin-isospin channels, with, possibly, different masses and
coupling constants, however, the momentum and energy dependence will be
the same. All the singularities are
treated a principal value integrals. The imaginary part, associated
with the homogeneous solution of the wave equation will follow
from the eigenvalue equation below.~\cite{Fan61}
The two different momentum dependencies are associated with two
different vertices (see Fig.~\ref{fig1}).
They serve as the two $\pi N\Delta$
couplings. Furthermore, there is a third coupling for the $\pi \pi N N$
contact interaction, which will yield the phase shift in the absence
of the $\Delta$ resonance in other spin-isospin channels. In lowest
order, the contact interaction is dominated by the pionic current
$[\pi^\ast, \partial_\mu \pi]$ which acts on the side of the 
triangle in Figure~\ref{fig1}. The two different time-orderings
will be associated with two different terms in the Hamiltonian.

For each channel and partial wave, in the rest frame, 
we can write down a general 
Hamiltonian for low-energy $\pi N$ scatttering based on the observations above:
\begin{eqnarray}
H & = & |\Delta\rangle m_\Delta \langle \Delta | + 
    \int_{m_\pi+m_N}^\infty d \epsilon |\pi N(\epsilon) \rangle \epsilon \langle \pi
    N(\epsilon)| \nonumber \\
& + & \sum_{n=0}^1 
\epsilon_{nk} 
\int_{m_\pi+m_N}^\infty d \epsilon' \sqrt{V_k(\epsilon')} |\pi N(\epsilon') \rangle 
       \int_{m_\pi+m_N}^\infty d \epsilon 
\langle \pi N(\epsilon) | \sqrt{V_n(\epsilon)} 
\nonumber \\
& + & \sum_{n=0}^1
\int_{m_\pi+m_N}^\infty d \epsilon | \Delta \rangle
\sqrt{W_n(\epsilon)}
\langle \pi N(\epsilon) |
                +  |\pi N(\epsilon) \rangle \sqrt{W_n(\epsilon)} \langle
\Delta | \ \ ,
\label{ham}
\end{eqnarray}
where $\epsilon_{nk}= 1 $ if $n\not = k$ and zero otherwise.
The momentum dependence of the contact interaction is based on 
a single $\pi \pi N N$ 
coupling through the pionic current $\sim [\pi,\partial_\mu \pi]$
with the nucleon current $\sim \bar N \gamma^5 \gamma_\mu N$,
where the single derivative acts either on the incoming or the
outgoing pion state, which leads to the asymmetric form.

The coupling functions $V$ and $W$, which correspond to local
interactions, contain both two terms with the same
functional form, however, different coupling constants ($V_n =
\frac{\lambda}{g_n} W_n$):
\begin{equation}
W_n(\epsilon) = g_n \frac{k^{2 n +2}}{\omega_N \omega_\pi}\frac{\partial
k}{\partial \epsilon} = \frac{g_n}{\epsilon}
\left( \frac{\sqrt{\epsilon^4 - 2 \epsilon^2 (m_\pi^2 +m_N^2) + 
(m_\pi^2 -m_N^2)^2}}{2\epsilon} \right)^{2
n + 1} \ \ ,
\end{equation} 
The kinematical factors $(\omega_\pi \omega_N)^{-1}$ make the
interaction relativistic invariant, and the factor $k^2$  for $n=1$
is due to the local, derivative coupling, in the rest frame. 
The coupling constants $g_n$ have a dimension $[{\rm
energy}^{2-2n}]$(GeV),
and $\lambda $ has a dimension of $[{\rm energy}^{1-2n}]$(GeV).
All the numerical factors from angular integrations, etc. are lumped in with
the coupling constants $g_n$ and $\lambda$.
The self-energy diagram, Eq.~(\ref{self}), is equivalent to ($g_n=1$):
\begin{equation}
\Sigma(E) = 
\int_{m_\pi+m_N}^\infty d \epsilon
\frac{[W_0(\epsilon);W_1(\epsilon)]}{E-\epsilon} \ \ ,
\end{equation}
which is again equivalent with the first order perturbation theory 
results for the Hamiltonian Eq.~(\ref{ham}).

This single channel problem with separable, but local interactions 
can be solved exactly, apart from the divergencies for which 
the finite renormalization has to be determined. 
The coupling functions $W$ and $V$ have the 
same functional form, therefore the ansatz wave function for the energy 
$\omega$ yields:
\begin{eqnarray} 
|\Psi(\omega)\rangle  & = & \alpha(\omega) |\Delta \rangle \\
& +  &\int
d \epsilon  \left( \frac{1}{\omega -\epsilon} - z(\omega) \delta (\omega -
\epsilon)\right)
(\sqrt{W_0(\epsilon)} \beta_0(\omega) +\sqrt{W_1(\epsilon)}
\beta_1(\omega)) |\pi N(\epsilon) \rangle \nonumber \\ 
& \equiv &
\alpha(\omega) |\Delta \rangle + \beta_0(\omega) |\omega(\pi N)_0 \rangle
+ \beta_1 (\omega) |\omega (\pi N)_1 \rangle  \ \ .
\end{eqnarray}
The function $z(\omega)$ models the ratio of the homogeneous and
the inhomogeneous part of the $\pi N$ scattering state. For this
single channel scattering state it is the only observable,
which is directly related to the phase shift.
When this wave function is inserted
into the continuum eigenvalue equation $0=(\omega -H)|\Psi(\omega)\rangle$
it yields the relations between the spectroscopic densities $\beta_i$ and 
the occupation number $\alpha$:
\begin{eqnarray}
 0 & = & \beta_0 |\omega(\pi N)_0 \rangle
+ \beta_1|\omega (\pi N)_1 \rangle  + (\omega - m_\Delta) 
\alpha |\Delta \rangle  \nonumber \\
& - & (\tilde W_0(\omega) - z W_0(\omega) ) \beta_0
| \Delta \rangle 
 -  (\tilde W_1(\omega) - z W_1(\omega) ) \beta_1
| \Delta \rangle \nonumber \\
& - & \alpha
(|\omega(\pi N)_0 \rangle +
|\omega(\pi N)_1 \rangle  )\nonumber \\
 & - & (\tilde V_0(\omega) -
 z V_0(\omega) ) \beta_0
|\omega(\pi N)_1 \rangle 
 -  (\tilde V_1(\omega) -
 z V_1(\omega) )\beta_1 
|\omega(\pi N)_0 \rangle \ \ ,
\label{M}
\end{eqnarray}
where the tilde refers to the Hilbert transform, without a factor of
$\pi^{-1}$:
\begin{equation}
\frac{g_n}{\lambda} \tilde V_n(\omega) = \tilde W_n(\omega) = 
\int d \epsilon \frac{W_n(\epsilon)}{\omega - \epsilon} \ \ .
\end{equation}
Projecting on the states $|\omega(\pi N)_0 \rangle$ and 
$|\omega(\pi N)_1 \rangle$ yields: $(i,j=0,1)$
\begin{equation} 
\beta_i  = \epsilon_{ij}\frac{1+ \tilde V_j(\omega) - z V_j(\omega)}{1 -  
[\tilde V_0(\omega) - z V_0(\omega)][\tilde V_1(\omega) - 
z V_1(\omega)]} \alpha \ \ ,
\end{equation}
inserting this back into Eq.~(\ref{M}) and projecting onto
$|\Delta \rangle$ yields a relation for $z$:
\begin{equation}
\omega - m_\Delta =  \frac{\sum_{i=0}^1 \epsilon_{ij} [\tilde
W_i(\omega) - z
W_i(\omega)][1+ \tilde V_j(\omega) - z V_j(\omega)]}{1 -
[\tilde V_0(\omega) - z V_0(\omega)][\tilde V_1(\omega) -
z V_1(\omega)]} \ \ ,
\label{zimp}
\end{equation}
which yields a straightforward but lengthy quadratic equation for $z$.
In the case of weak contact interactions $\lambda \sim 0$, the solution can
be approximated by:
\begin{equation}
z = -\frac{1}{W_0 + W_1}
( \omega - m_\Delta - \tilde W_0 - \tilde W_1 ) \ \ ,
\label{z}
\end{equation}
which can be cast in the form of an analyticity corrected 
energy-dependent Breit-Wigner resonance. 

The general phase shift is given by:
\begin{equation}
\tan \delta = \frac{\pi}{z} \ \ .
\end{equation}
Since there is only one channel, the scattering process is fully
characterized by $z$ through the phase shift. Only the Hilbert
transforms
of $W_0 = \frac{g_0}{\lambda} V_0 $ and $W_1 = \frac{g_1}{\lambda} V_1$
still need to be determined.

The backward or $Z$-diagram contributions are included, to make
the results covariant and the expressions simpler. The singularity
is treated as a principal value singularity:
\begin{equation}
\tilde W_{n}(\omega) = g_n \int \frac{d s }{\sqrt{s}}
\left(\frac{\sqrt{s^2 - 2 s (m_\pi^2 + m_N^2) + (m_\pi^2 -
m_N^2)^2}}{2 \sqrt{s}} \right)^{2 n+1} \frac{1}{\omega^2 - s} \ \ .
\end{equation}
The integral is divergent.
Subtractions make the integrals finite:
\begin{equation}
\tilde W^{\rm r}_n(\omega) = \int \frac{\omega^{2n+2}}{s^{n+1}} d s
W_n(\sqrt{s}) \frac{1}{\omega^2 - s}  \ \ ,
\end{equation}
which yield an analytical results for the Hilbert transform
for an arbitrary integer value of $n$:
\begin{equation}
\tilde W^{\rm r}_{n}(\omega) = \frac{-g_n (1 -
{\cal T}_{2n +1})}{\omega (2 \omega)^{2n+1}} 
{\rm Re} \left[\left(\sqrt{A}
\right)^{2 n+1} \log \left(
\frac{ - \omega^2 + m_\pi^2 + m_N^2 + \sqrt{A}}{m_\pi m_N} \right)
\right] \ \ ,
\end{equation}
where ${\cal T}_m$ stands for the $m$-th order Taylor expansion in
$\omega^2$ around $\omega^2 = 0$, and 
\begin{equation} 
\sqrt{A} \equiv
\sqrt{\omega^4 - 2 \omega^2 (m_\pi^2 + m_N^2) +(m_\pi^2 - m_N^2)^2} \ \ .
\end{equation}
It should noted that the real part of the function above arises through
the real part of the logarithm and the real part of the squareroot for
$\omega^2 >(m_N + m_\pi)^2$ and $\omega^2 <(m_N - m_\pi)^2$ and through the 
product of the imaginary parts of both for 
$(m_N - m_\pi)^2<\omega^2<(m_N + m_\pi)^2$.

The subtractions correspond to two divergent constants that should
be reabsorbed in the mass and the wave function of the bare
$\Delta$ state:
\begin{eqnarray}
\tilde W_0(\omega) & =  & \tilde W^{\rm r}_0(\omega) + c^{(0)}_0 \ \ , \\
\tilde W_1(\omega) & =  & \tilde W^{\rm r}_1(\omega) + c^{(1)}_0 +
c^{(1)}_1 \omega^2 \ \ ,
\end{eqnarray}
where the renormalization constants $c$ are fixed through:
$\tilde W_n(m_\Delta) = (\omega^2 - m_\Delta^2)^{n} + {\cal
O}(n+1)$. Arguably, $W_0$ could have a finite wave function
renormalization. 
However, as one can see from Eq.~(\ref{z}) this renormalization
factor $c^{(0)}_1$ can be reabsorbed in an overall shift of the coupling
constants $g_n \to g_n/c$, for the only observable quantity
$z$, where $\omega^2 - m_\Delta^2$ is approximated by 
$2 m_\Delta (\omega - m_\Delta)$. The coupling functions $W$
can clearly be identified with the bare $\Delta $ state; the constant
part of $\tilde W$ is indistinquishable from the mass $m_\Delta$. For the
contact interaction there is no such identification, therefore
these renormalization constants will be determined from the data.

\begin{figure}
\centerline{\includegraphics[width=8cm]{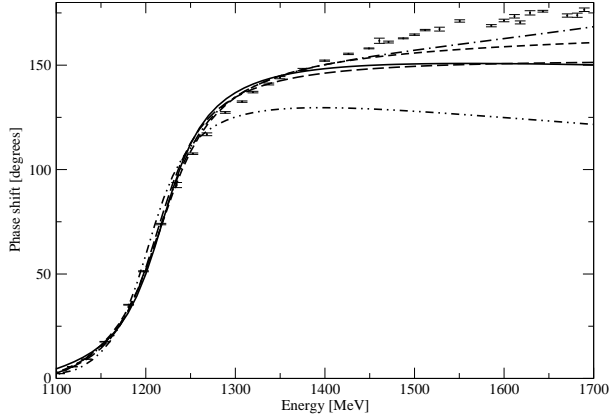}}
\caption{The phase shift in the $P_{33}$ scattering channel
which contains the $\Delta$ resonance. The dashed-double-dotted
line is a fit with a dynamically generated mass through the
contact interaction with $\lambda$. The solid line is a
fit with $g_0$ and $g_1$. The dashed line is a fit with
$g_0$, $g_1$, and $\lambda$. The long-dashed line is a
fit with $g_\frac{1}{2}$, which has a much better $\chi$-square
value due to a better threshold behavior. The dot-dashed line 
is a fit with two $g_\frac{1}{2}$ couplings to two different 
resonances. } \label{fig2}
\end{figure}

\section{Without bare resonance state}

In the case of only contact interactions and no resonance state the
expressions simplify somewhat.  It corresponds to the Hamiltonian,
Eq.~(\ref{ham}), with $W_0=0$ and $W_1=0$, which decouples the
resonance from the scattering states.
The ratio of the homogeneous to inhomogeneous solution $\pi/z(\omega)$
that leads to the phase shift is given by:
\begin{equation}
z = \frac{V_0 \tilde V_1 + V_1 \tilde
V_0 + \sqrt{(V_0 \tilde V_1 + V_1
\tilde V_0)^2 + 4 V_0  V_1 (1- \tilde V_0 
\tilde V_1)}}{2  V_0  V_1} \ \ .
\end{equation}
Depending on the coupling constant $\lambda$, the phase shift can be positive
or negative. In the physical range $m_N + m_\pi < \omega< 2.0$ GeV the
sign change in the phase shift occurs for $\lambda$ varying between 1.0 and 3.0. 
In that case the finite renormalization is set to zero, such that $V_i$ 
vanishes at $\omega = 0$.

It is possible to mimic part of the resonance with a mass $m_V$
through a non-trivial
renormalization of $\tilde V_0$ and $\tilde V_1$:
\begin{eqnarray}
\tilde V_0 & = & \tilde V_0^{\rm r} + c^{(\lambda,0)}_0 \\
\tilde V_1 & = & \tilde V_1^{\rm r} + c^{(\lambda,1)}_0  + 
c^{(\lambda,1)}_1 \omega^2 \ \ ,
\end{eqnarray}
where $m^2_V =  -(\tilde V_1^{\rm r}(m_V) +
c^{(\lambda,1)}_0)/(c^{(\lambda,1)}_1 \omega^2 + \partial_{\omega^2}\tilde V_1^{\rm
r}(m_V))$.
This is normally referred to as a dynamically generated resonance.

\section{Results}

In order to test the quality of the model a whole set of fits were
performed, excluding certain parts of the interaction, and including others.
The least-square fit is determined by minimizing the $\chi$-square value.
However, since there is a definite non-linear dependence of the
scattering data on the coupling constants some care is required. As 
expected already above, fitting with both the contact interaction
and the bare resonance is indeterminate, as both yield a similar effect
and a flat minimum in parameter space for $\chi$-square.

However, as it turns out all the possible fits for the $\Delta$
resonance in the $P_{33}$ channel do much worse that a
fit with a quasi-local interaction $n=\frac{1}{2}$. This interaction
has the wrong dimensionality, however, produces the best fit with
two parameters: the coupling constant and the renormalized resonance
mass. This is also the only way to yield the right threshold behavior,
if the fit includes data points beyond the resonance.

The various results are listed below and shown in Fig.~\ref{fig2}. 
In the cases that the bare
resonance is presence, i.e., $g_0 \not = 0$ or $g_1 \not = 0$, 
its mass is set to zero, i.e., $m_\Delta =0$, 
the renormalization constants $c_0$ will account for the mass, with 
a finite renormalization. The energy for which $z=0$ and the phase shift
crosses $90^\circ$ is for every fit within a few
MeV of the $\Delta $ resonance at $1.232$ GeV. The number of data points
for the fits ranges between 30 and 50 depending on the range.

\begin{figure}
\centerline{\includegraphics[width=8cm]{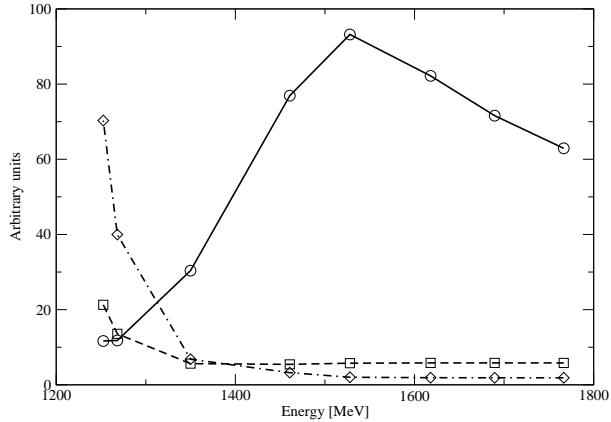}}
\caption{The quality of the fit and the variation of the
coupling constants for the $P_{33}$ channel. The fit 
with $g_0$ and $g_1$
includes all the data points from threshold to the energy
indicated. The circles are the $\chi^2/\#$ , the squares
the values of $g_0$ in $100 {\rm GeV}$, and the diamonds are the
values of $g_1$ in $2 {\rm GeV}$. The lines only guide the
eye. The $\chi^2/\#$ drops for large energies due to
the poorer quality of the data at higher energies. It is clear
from the fits that $g_1$ dominates at threshold, however,
cannot fit the data alone at higher energies. There are
three parameters fitted: the $\Delta$ mass and the coupling
constants $g_0$ and $g_1$,
the smallest number of data points is nine.} \label{fig2b}
\end{figure}

\begin{figure}
\centerline{\includegraphics[width=8cm]{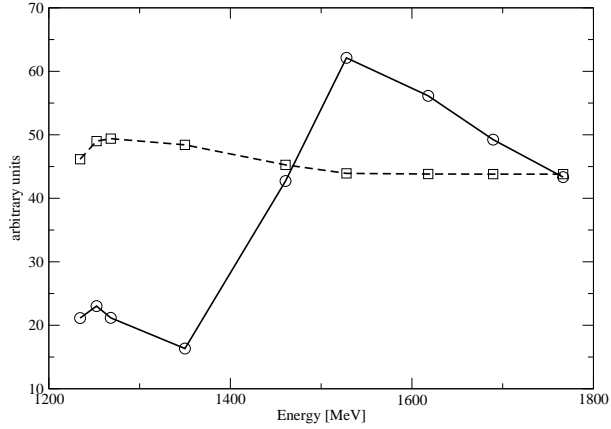}}
\caption{The quality of the fit with the pseudo-local
coupling $g_\frac{1}{2}$. The circles are the $\chi^2/\#$
values. The squares are the values of the coupling constants
in $100 {\rm GeV}$.} \label{fig2c}
\end{figure}

\begin{itemize}
\item{For the range 1.1 GeV to 1.6 GeV, fitting only with nonzero
$g_0$ yields $\chi^2/\#= 120$ and $g_0 = 0.082$ and $c_0^{(0)} = 1.294 $.}
\item{For the range 1.1 GeV to 1.6 GeV, fitting only with nonzero 
$g_1$ yields $\chi^2/\#= 208$ and $g_1 = 76000.$ and
$c_0^{(1)} = 10100. $. The results do not depend significantly on
$c_1^{(1)}$, which is subsequently set to zero.
However, if we set $c_1^{(1)}=1$, we find more appropriate numbers
for $g_1 =  3.78$ and $c_0^{(1)} = 0.493$, since for this
particular energy $\partial_\omega W^{\rm r}_1(m_\Delta)$ gives a 
finite wave-function renormalization near unity, which makes the
energy dependence of $z$ very weak.}
\item{For the range 1.1 GeV to 1.6 GeV, fitting only with nonzero 
$\lambda$ yields $\chi^2/\#= 140$, $\lambda = 2.31$, $c^{(\lambda,1)}_0 = 2.26$,
$c^{(\lambda,1)}_1 = -1.47$, and $c^{(\lambda,0)}_0 = 8.69$. The fit
yields a better threshold behavior but a worse high-energy behavior
compared to the $g_0+g_1$ fit.} 
\item{For the range 1.1 GeV to 1.6 GeV, fitting  with nonzero $g_0$
and $g_1$ yields  $\chi^2/\#= 82$, $g_0 = 0.060$, $g_1 = 0.79$,
$c_1^{(0)} = 0.69$, and $c_1^{(1)} = 0.69$. Both renormalization 
constants take about half the resonance mass, changing this fraction
does not change the $\chi^2$ value. Changing the range of the
data set changes the results substantially, as can be seen in 
Figure~\ref{fig2b}. At threshold till 1.25 GeV the $g_1$ coupling
fits the data well.}
\item{For the range 1.1 GeV to 1.6 GeV, fitting only with nonzero 
$g_\frac{1}{2}$ yields  $\chi^2/\#= 56$, $g_\frac{1}{2} = 0.44$, and
$c^{(\frac{1}{2})}_0 =  1.227 $. It is by far the best fit
archieved.}
\item{For the extended range 1.1 GeV to 1.8 GeV, the fit even
improves, due to poorer quality data: $\chi^2/\#= 45$ and 
$g_\frac{1}{2} = 0.44$, and $c^{(\frac{1}{2})}_0 =  1.227$. The actual
values of the coupling constant and the renormalization constant
do not change. For the whole range the two parameters $g_\frac{1}{2}$
and $m_\Delta$ fit the data with a consistent coupling strength of
$g_\frac{1}{2} = 47 \pm 2$, as can be seen in Figure~\ref{fig2c}. Fits
with data below
the inelastic threshold the $\chi^2/\#$ ranges between 16 and 23.}
\item{A further dramatic improvement is archieved with an additional
resonance. For the range  1.1 GeV to 2.0 GeV, the fit yields:
$\chi^2/\#= 13$, $g_{a\frac{1}{2}} = 0.48$, $c^{(a,\frac{1}{2})}_0 =  
1.231$, $g_{b\frac{1}{2}} = 1.75$, and
$c^{(b,\frac{1}{2})}_0 = 3.74$. The second resonance is around 2.5-4.0 GeV,
however, its position cannot be determined accurately from the data,
since it lies outside the range.}
\end{itemize}

\begin{figure}
\centerline{\includegraphics[width=8cm]{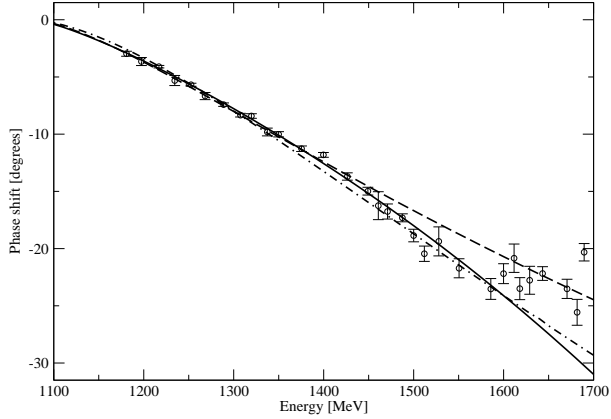}}
\caption{The phase shift in the $P_{31}$ scattering channel.
The long-dashed line is a fit with $g_1$ only. The dot-dashed
line is a fit with $\lambda$ only. The solid line is a
fit with $g_0$, $g_1$, and $\lambda$, which does not yield a
significant improvement at low energies.}\label{fig3}
\end{figure}

\begin{figure}
\centerline{\includegraphics[width=8cm]{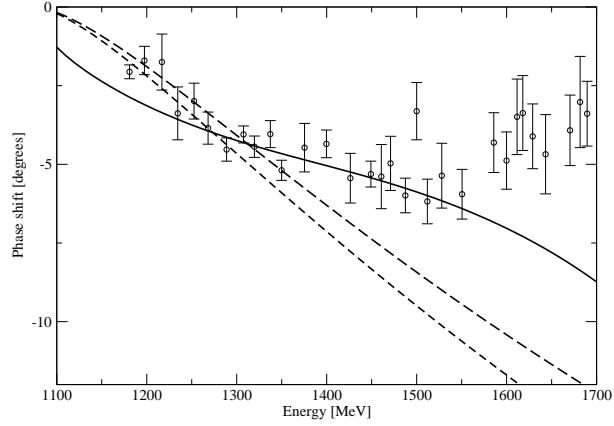}}
\caption{The phase shift in the $P_{13}$ scattering channel.
The dashed line is a fit with $g_1$ only. The long-dashed
line is a fit with $\lambda$ only. The solid line is a
fit with $g_0$, $g_1$, $\lambda$. }\label{fig4}
\end{figure}

Although, consistency with non-relativistic treatment, where the
power $n$ in $W_n$ is directly related to the partial wave, of the
$\Delta$
resonance would require that $g_0$ is zero, it is impossible
to fit the $P_{33}$ data properly with that assumption. Attempts
to make sensible fits with $g_1$ and two bare states failed. The
accurate data at low energies forces a $g_\frac{1}{2}$ interaction
in the case of two bare states.  However, the
$P_{31}$ and the $P_{13}$ channels do give results consistent
with $g_0 = 0$. A number of fits were made, which are listed below,
and show respectively in Fig.~\ref{fig3} and Fig.~\ref{fig4}.

\begin{itemize}
\item{For the $P_{31}$ a fit with $g_1$ only gives excellent results
from the threshold up to the first structure in the data at 1.55 GeV.
The $\chi$-square is comparable to unity.}
\item{Similarly for the $P_{31}$ channel, a fit with $\lambda$ only
gives good results too. However, the second solution of the quadratic equation
starts to interfere with the physical solution at 1.6 GeV.}
\item{Combining $g_1$ and $\lambda$ interactions does not give
a significant improvement with respect to either fit.}
\item{For the $P_{13}$ the data is rather coarse. Some structure appears
at 1.4 GeV. Both $g_1$ on its own, as $\lambda$ on its own give a fit
with a $\chi$-square comparable to unity. The shape of the fit depends
on the range of the data points, however, as the $\chi$-square
is comparable to unity, no conclusions can be drawn.}
\item{Combining both  fits for the $P_{13}$ data leads to an
improvement as some of the structure at 1.7 GeV is reproduced, however,
as the threshold behavior fails, it is considered an artificial improvement.}
\end{itemize}

Since in the case of the $P_{31}$ and the $P_{13}$ channels there is
no resonance that fixed a scale, there is considerable freedom to
scale the coupling constants and renormalization constants and still
produce the same results. However, especially for a non-zero $\lambda$,
there seems to be no obvious rebundant parameters in the model, wich can 
be fixed a priori. The different coupling constants and renormalization
constants are not quoted as they, on their own, have lost their significance.

\section{Conclusions}

A local field theory without cut-offs or form factors should fit the data
at all energies. Deviations can only occur if other decay channels
open, or higher-order terms in the interaction are important.
Therefore we expect the data to be fitted well with our model
below the two-pion production threshold, which is the case for
the $P_{31}$ and the $P_{13}$ channels. In the same theory resonance
states can be pushed out of the physical energy region, and
resonances can be generated dynamically through a contact interaction.
Eventually it is the quality of the fit that determines which is
the likely process. For the $P_{33}$ channel the bare state gives
a better results, for the $P_{31}$ and the $P_{13}$ channels both
the contact interaction and the bare resonance state whose pole
is pushed away from the real axis through a strong interaaction do
a good job fitting the data below 1.5 GeV.

From the results it is clear that one cannot simply add a resonance
structure to a contact interaction incoherently. The effects of the
contact interaction are not like what one would expect from perturbation
theory. When the interaction is summed to all orders, the resonance 
dominate the results. Any other, perturbative, effects will be 
irrelevant for the phase shift around the resonance peak. In all the 
different versions of the model, where some interactions
are switched off, it is impossible to create the effect of a rising
phase shift beyond $180^\circ$ after the $\Delta$ resonance, 
usually associated with
a background, without distorting the resonance itself. It can only 
consistently be explained by a second 
resonance, which is found around 2.5-4.0 GeV.

Furthermore, from the data the $W_\frac{1}{2}$ coupling function
prevails over the $W_0$ and the $W_1$ coupling functions, or even
a combination of both with five fitting parameters, compared to
the two of the $W_\frac{1}{2}$ coupling alone.

The same model, with different values for the coupling constants
and the renormalization constants, fit the low-energy $P_{13}$ 
and $P_{31}$ data well, which have no resonance structure. 
The fits give a model in with the derivative
coupling dominates, associated with the $g_1$ coupling constant.

\end{document}